\RequirePackage{fix-cm}
\documentclass[3p,twocolumn,number,sort&compress]{elsarticle}
\usepackage{graphicx}
\usepackage{amsmath}
\usepackage{upgreek}
\usepackage{color}
\usepackage{siunitx}
\usepackage{a4wide}
\usepackage{fancyhdr}

\DeclareSIUnit\atomicmassunit{amu}
\DeclareSIUnit\ion{ion}
\makeatletter
\def\ps@pprintTitle{%
	\let\@oddhead\@empty
	\let\@evenhead\@empty
	\def\@oddfoot{}%
	\let\@evenfoot\@oddfoot}
\makeatother

\begin{document}

\begin{frontmatter}
\title{Development of a recoil ion source providing slow Th ions including \textsuperscript{229(m)}Th in a broad charge state distribution}


\author[1,2,3,4]{Raphael Haas}
\author[1,2]{Tom Kieck}
\author[2,4,5]{Dmitry Budker}
\author[1,2,3,4]{Christoph E. D\"ullmann}
\author[6]{Karin Groot-Berning}
\author[2,4]{Wenbing Li}
\author[1,2]{Dennis Renisch}
\author[2,6]{Ferdinand Schmidt-Kaler}
\author[6]{Felix Stopp}
\author[2,6]{Anna Viatkina}

\address[1]{Department Chemie, Johannes Gutenberg-Universit\"at Mainz, 55128 Mainz, Germany}
\address[2]{Helmholtz-Institut Mainz, 55099 Mainz, Germany}
\address[3]{GSI Helmholtzzentrum für Schwerionenforschung GmbH, 64291 Darmstadt, Germany}
\address[4]{PRISMA Cluster of Excellence, Johannes Gutenberg-Universit\"at Mainz, 55128 Mainz, Germany}
\address[5]{Department of Physics, University of California, Berkeley, California 94720-7300, USA}
\address[6]{QUANTUM, Institut f\"ur Physik, Johannes Gutenberg-Universit\"at Mainz, 55128 Mainz, Germany}

\begin{abstract}
	Ions of the isomer \textsuperscript{229m}Th are a topic of high interest for the construction of a ``nuclear clock" and in the field of fundamental physics for testing symmetries of nature. They can be efficiently captured in Paul traps which are ideal for performing high precision quantum logic spectroscopy. Trapping and identification of long-lived \textsuperscript{232}Th\textsuperscript{+} ions from a laser ablation source was already demonstrated by the TACTICa collaboration on Trapping And Cooling of Thorium Ions with Calcium. The \textsuperscript{229m}Th is most easily accessible as $\alpha$-decay daughter of the decay of \textsuperscript{233}U. We report on the development of a source for slow Th ions, including \textsuperscript{229(m)}Th for the TACTICa experiment. The \textsuperscript{229(m)}Th source is currently under construction and comprises a \textsuperscript{233}U monolayer, from which \textsuperscript{229(m)}Th ions recoil. These are decelerated in an electric field. Conservation of the full initial charge state distribution of the \textsuperscript{229(m)}Th recoil ions is one of the unique features of this source. We present ion-flight simulations for our adopted layout and give a final source design. This source will provide Th ions in their original charge state at energies suitable for capture in a linear Paul trap for spectroscopy investigations.
\end{abstract}
\end{frontmatter}
\section{Introduction}
\label{intro}
The \textsuperscript{229m}Th became a nuclide of high interest in the field of quantum physics in recent years because of its low-energy isomeric nuclear state \cite{Safronova2018a}. This is considered a promising candidate for the development of a nuclear clock \cite{Peik2003} and is an interesting nuclide in the search for physics beyond the standard model \cite{Flambaum2006}. Basic features of the \textsuperscript{229m}Th were experimentally studied in recent years \cite{Wense2016,Seiferle2017,Thielking2018,Seiferle2019}. Its low excitation energy \cite{Seiferle2019} in principle allows direct optical excitation with existing laser technology, but the excitation energy is presently not known with sufficient accuracy. Recently, the isomeric state of \textsuperscript{229}Th was successfully populated in the $\beta$\textsuperscript{-}-decay of the artificial nuclide \textsuperscript{229}Ac \cite{Shigekawa2019} and via synchrotron X-ray pumping to the second excited state at \SI{29.19}{\kilo\electronvolt} \cite{Masuda2019}. At the moment, the simplest experimental access to \textsuperscript{229m}Th is via $\alpha$-decay of its mother nuclide \textsuperscript{233}U, which proceeds through the isomeric state in \SI{2}{\percent} of all decays \cite{Thielking2018,Barci2003}. The aim of the collaboration TACTICa\footnote{TACTICa (\underline{T}rapping \underline{A}nd \underline{C}ooling of \underline{T}horium \underline{I}ons with \underline{Ca}lcium)} is to capture \textsuperscript{229(m)}Th and further Th isotopes inside a \textsuperscript{40}Ca\textsuperscript{+} Coulomb ion crystal by sympathetic laser cooling in a Paul trap for precision spectroscopy experiments. This has successfully been demonstrated with \textsuperscript{232}Th\textsuperscript{+} ions from a laser ablation source \cite{Groot-Berning2019,Stopp2019}. There, the \textsuperscript{232}Th\textsuperscript{+} ions became a high mass defect inside the \textsuperscript{40}Ca\textsuperscript{+} Coulomb crystals \cite{Stopp2019}. The used segmented Paul trap is capable of capturing singly-charged ions with kinetic energies of up to \SI{1}{\kilo\electronvolt}. It operates at a maximum background pressure of \SI{1E-10}{\milli\bar}. Owing to these specifications, the loading of \textsuperscript{229}Th recoil ions from the decay of \textsuperscript{233}U with a kinetic energy of \textit{E}\textsubscript{k} = \SI{84}{\kilo\electronvolt} [derived from \textit{E}\textsubscript{k}(\textsuperscript{229}Th) = \textit{Q}\textsubscript{$\alpha$}(\textsuperscript{233}U) - \textit{E}\textsubscript{$\alpha$}(\textsuperscript{233}U), where \textit{Q}\textsubscript{$\alpha$} is the Q-value of the $\alpha$-decay and \textit{E}\textsubscript{$\alpha$} the kinetic energy of the $\alpha$-particle] becomes challenging. Additionally, the ions emerge from the sample in a rather wide charge-state distribution. This was studied in \textsuperscript{222}Ra from \textsuperscript{226}Th \cite{Gunter1966} and is isotope-specific, as it depends on the nuclear structure and transitions of the individual nuclei. Whereas in \cite{Wense2016,Seiferle2017,Thielking2018,Seiferle2019}, \textsuperscript{229}Th\textsuperscript{3+} has been successfully extracted as low-energy ion beam from a buffer-gas-stopping-cell based setup, this approach cannot be used for deceleration in our setup due to the gas load, which is incompatible with the pressure requirements in the TACTICa Paul trap. Additionally, this technique leads to a loss of the higher charge states, which are of interest for spectroscopic investigations of the isomeric state. There are good reasons to perform atomic spectroscopy of highly charged ions of Th. To start with, little experimental information is available about the energy levels of Th for ionization stages higher that 2+ \cite{Kramida2019}, even though recent calculation were carried out for Th\textsuperscript{4+} \cite{Safronova2018b}. Apart from serving as a test of atomic theory, spectroscopy in highly charged states of Th may enable studies directly relevant to the isomeric nuclear excitation via processes such as the so-called “electron bridge” (see, e. g., \cite{Mueller2017}, which describes this process in 3+ ions). Therefore, a novel ion source has been developed, which will provide low-energy \textsuperscript{229m}Th\textsuperscript{n+} ions produced as recoil daughter products in the $\alpha$-decay of \textsuperscript{233}U. The concept of the source is based on electrostatic deceleration of ions from the \textsuperscript{233}U recoil ion source. It will consist of a monolayer sample of \textsuperscript{233}U to provide a monochromatic \textsuperscript{229(m)}Th\textsuperscript{n+} beam. The source will be set to a high negative potential to decelerate the recoil ions emitted from the source surface when running up against ground potential, at which the vacuum chamber is kept. The absence of gas encounters provides the opportunity to study \textsuperscript{229(m)}Th in various charge states like in the case of \textsuperscript{222}Ra from \textsuperscript{226}Th-decay \cite{Gunter1966}. In addition to the recoil ion source, a laser-ablation ion source \cite{Groot-Berning2019} will also be installed to provide ions of Th isotopes that are available as macroscopic samples. The combination of the ablation and recoil ion sources offers the possibility to study a variety of Th isotopes (see Fig. 1) in several charge states. The ions will be captured by a Paul trap, sympathetically cooled using a crystal of laser-cooled Ca\textsuperscript{+} ions and be available for precision spectroscopy. This is of interest for the verification of hypothetical extensions of the theoretical description of the isotope shift using the nonlinearity of the King plot and thus for the search for new particles \cite{Flambaum2018}.
\begin{figure*}
	\centering
	\includegraphics[width=1.0\textwidth]{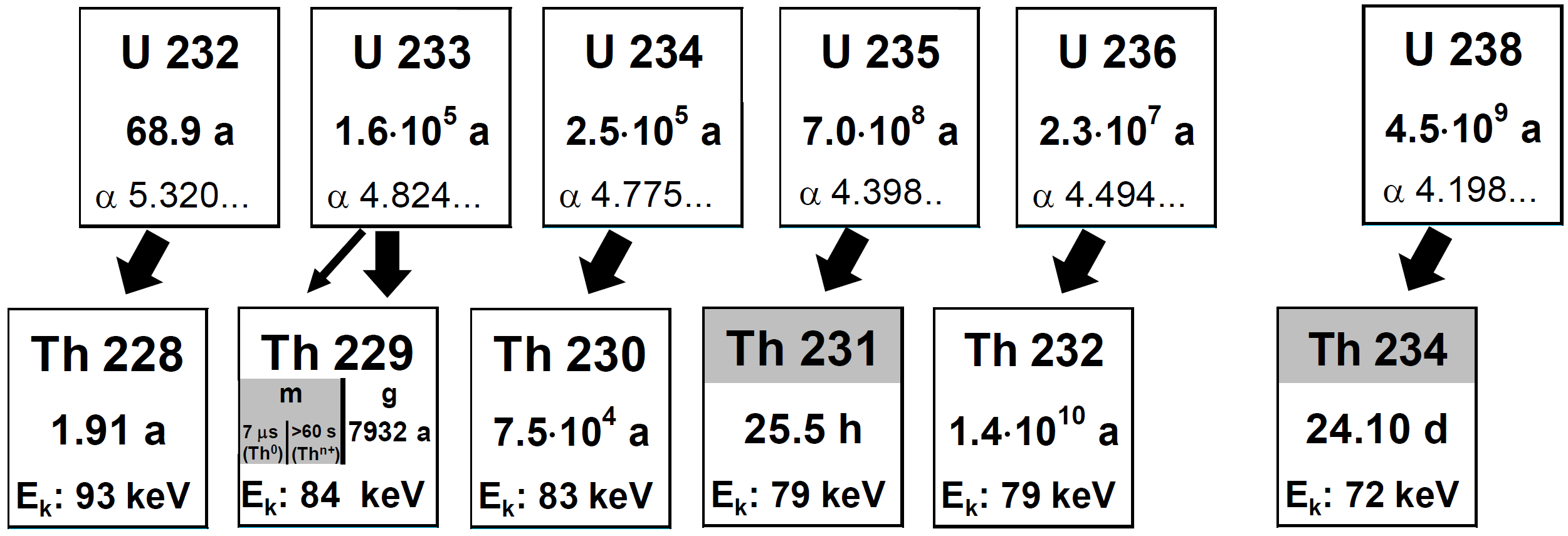}
	\caption{Excerpt of the nuclear chart showing thorium isotopes of interest for TACTICa. Long-lived uranium alpha-decay precursor isotopes are also shown. The half-lives of all isotopes are indicated. For the uranium isotopes, the $\alpha$-particle energies are given in \SI{}{\mega\electronvolt}. For the thorium isotopes, the kinetic energies, E\textsubscript{k}, of the recoil ions populated in the $\alpha$-decay are given. Many of the thorium isotopes are available in macroscopic quantity, suitable for ablation source production. Those indicated in gray are only available as $\alpha$-decay products from the recoil ion source.}
	\label{fig:1}
\end{figure*}

\section{Electrostatic deceleration}
\label{sec:1}
In the following, we use \textsuperscript{229(m)}Th as the example nuclide. The kinetic energy of the \textsuperscript{229(m)}Th recoil ions of about \SI{84}{\kilo\electronvolt} has to be reduced before the ions enter the Paul trap to allow their capture. As mentioned in section \ref{intro}, the maximum acceptable kinetic energy of singly-charged \textsuperscript{229(m)}Th ions is around \SI{1}{\kilo\electronvolt}. A recoil ion source provides \textsuperscript{229(m)}Th recoil ions in a  2$\pi$ angular distribution. Ion cooling provides both the reduction of the transverse component of the ions' velocity and longitudinal deceleration of the recoil ions. This would be the ideal solution for a 2$\pi$ source to produce an ion beam. The most efficient cooling method is buffer gas cooling as used in other \textsuperscript{229(m)}Th experiments \cite{Wense2016,Seiferle2017,Thielking2018,Seiferle2019}. Owing to charge exchange collisions in a buffer gas cell, the charge state distribution collapses to low charge states. In \cite{Wense2016,Seiferle2017,Thielking2018,Seiferle2019}, predominantly 2+ and 3+ ions were extracted. To retain higher charge states, which are expected to be populated in the $\alpha$-decay of \textsuperscript{233}U, gas-free cooling methods are required.

We decided to use electrostatic deceleration of the recoil ions by a negative potential on the source. For this approach, a source with a high recoil efficiency as well as a sharp recoil energy spectrum without any loss of the high charge states is needed. This is typically given by an ideal recoil source consisting of a single atomic layer of \textsuperscript{233}U. Such sources can be efficiently produced by several methods, e. g., at the Institute of Nuclear Chemistry in Mainz. A comparison of the different source fabrication methods will be published separately \cite{RHaas}.
The basic idea of the electrostatic deceleration of recoil ions in the TACTICa source is to use the potential difference between the source (kept at negative voltage) and the trap entrance (at ground potential). Because of the kinetic energy of the recoil ions of about \SI{84}{\kilo\electronvolt}, a high negative potential between \SIrange{83}{84}{\kilo\electronvolt} is needed to decelerate singly-charged \textsuperscript{229}Th\textsuperscript{+} ions to an energy below \SI{1}{\kilo\electronvolt}. For higher charge states, correspondingly lower voltages are sufficient. At a specific voltage, the source will deliver the chosen charge state with proper energy as well as lower charge states with higher energy due to lower deceleration. Higher charge states will always be suppressed.

\begin{figure*}
	\centering
	\includegraphics[width=1.0\textwidth]{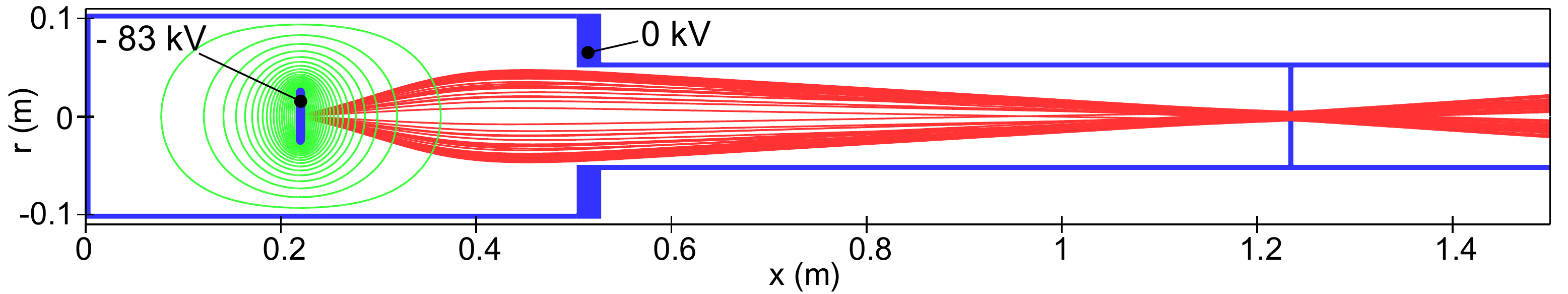}
	\caption{Flight trajectories of the singly-charged recoil ions that clear the aperture and equipotential lines. The ion energy is about \SI{1000}{\electronvolt} after exiting the electrostatic deceleration chamber.}
	\label{fig:2}
\end{figure*}

The advantage of this source type concerning charge-state conservation comes with the disadvantage of low efficiency, because of low angular acceptance caused by emittance conservation (Liouville's theorem). \SI{50}{\percent} of all ions are emitted into the substrate and therefore lost. The emittance of the residual ions is preserved regardless of ion optics and could only be reduced by cooling mechanisms. Many of these (e.g., buffer gas \cite{Wense2016}, stochastic \cite{Moehl1980}, electron \cite{Budker1976}, laser \cite{Schroeder1990}) are not applicable here. In our design, the source is kept at high voltage matching the charge state and only the vacuum chamber at ground potential acts as an ion-optical element. In our simulation model used for the optimizations of the source, the ions are emitted from a round surface with a diameter of \SI{30}{mm} in 2$\pi$ angular distribution with an energy of \SI{84}{\kilo\electronvolt} randomized in position and angle with the Mersenne Twister 19937 algorithm \cite{Matsumoto1998}. Several ion-flight simulations were performed with the open-source software IBSimu \cite{Kalvas2010} to evaluate the transmission efficiency of the recoil ions through the first aperture with a diameter of \SI{100}{mm}. The simulation setup including solid ion-optics, equipotential lines and flight pathways of the aperture-transmitting ions is given in Fig. \ref{fig:2}.

\section{Final design}
\label{sec:2}
\begin{figure*}
	\centering
	\includegraphics[width=1.0\textwidth]{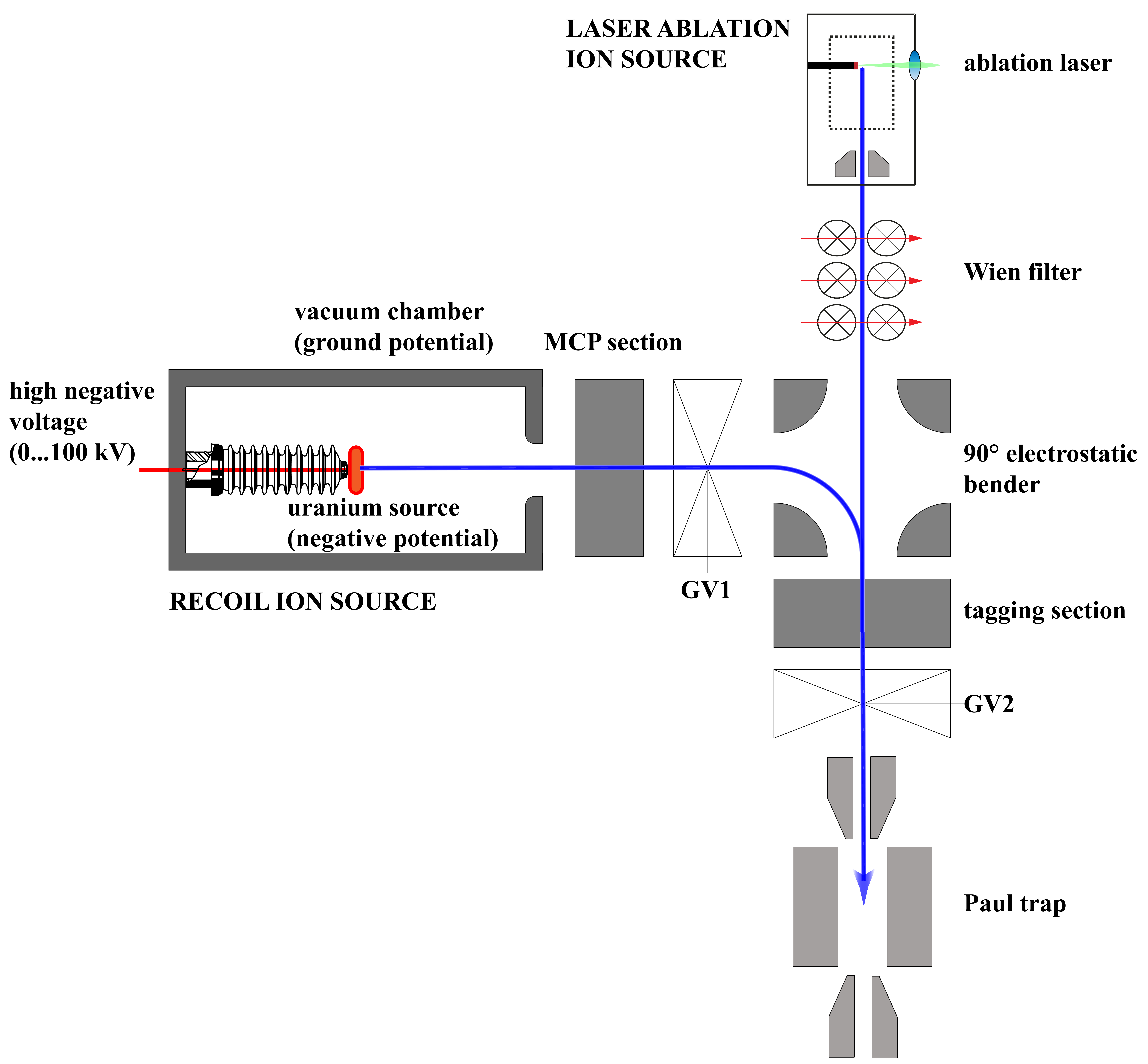}
	\caption{Schematic illustration of the TACTICa experimental setup including both the recoil ion source and the laser ablation ion source (not to scale). See text for details.}
	\label{fig:3}
\end{figure*}

The final design of the TACTICa source is kept simple and is shown schematically in Fig. \ref{fig:3}. As mentioned in section 2, the electrostatic deceleration of the recoil ions can be achieved with a high negative potential on the source and the vacuum chamber at ground potential. The \textsuperscript{233}U source is fixed in the center of a tubular chamber of a \SI{500}{mm} length and an inner diameter of \SI{200}{mm}. After deceleration at the end of the tubular chamber, the ions enter a section with a microchannel plate detector (MCP) for beam-analysis. The source chamber as well as the beam analysis chamber can be separated from the main chamber containing the trap with a gate valve (GV1) for maintenance and source exchange. A \SI{90}{\degree} electrostatic quadrupole bender is the connection point of both the recoil ion source and the laser ablation ion source with the trap. It is used to switch between the ion beams of both sources. In the ablation ion source, singly-charged ions are produced by means of a pulsed laser. The laser pulses serve as time-stamps, leading to precisely known ion arrival times at the Paul trap, which can therefore be closed at times synchronized with incoming ions, thus ensuring the ions' capture \cite{Groot-Berning2019}. To prevent ions other than Th from entering the trap, a Wien filter is used, which selects the right mass-to-charge ratio. In case of the recoil ion source, ion production by $\alpha$-decay of the mother nuclide occurs at a random time, and no timing information is available in this case. Therefore, a small chamber housing a tagging section to obtain timing information of passing recoil ions is placed in front of the Paul trap. This will provide the signal necessary to know when exactly the trap needs to be closed. The tagging methods under investigations are by image charge detection \cite{Raecke2018}, by fluorescence detection \cite{GarciaRuiz2017} and by energy loss in the ion crystal \cite{Schmoeger2015} followed by doppler-recooling measurement \cite{Wesenberg2007,Huber2008}. The section with the ablation ion source, the quadrupole bender and the tagging station can also be separated for maintenance from the Paul trap with a gate valve (GV2), to keep the trap under ultra-high vacuum conditions also when the vacuum has to be broken outside of the trap.

\section{Conclusion}
\label{conclusion}
We present the design of a recoil ion source setup using electrostatic deceleration for the TACTICa experiment. This source will deliver Th\textsuperscript{n+} recoil ions of a variety of Th isotopes (including \textsuperscript{229(m)}Th) from a uranium monolayer source. The kinetic energy for trapping of the ions has to be below \SI{1}{\kilo\electronvolt}. The most promising way to decelerate the ions and to retain them in their original charge states is via electrostatic deceleration. It allows to select a single charge state of the full initial charge state distribution \cite{Gunter1966}. Furthermore, electrostatic bending removes expected contamination of \textsuperscript{228}Th ions due to their higher kinetic energy of \SI{93}{\kilo\electronvolt}. Detailed simulations with IBSimu \cite{Kalvas2010} are ongoing to optimize the setup with respect to the transmission rate. The rate of the most interesting ion \textsuperscript{229m}Th will be nevertheless low, as it is restricted by both the activity of approximately \SI{13}{\becquerel\per\square\centi\meter} determined by the source size and half-life of \textsuperscript{233}U and a population rate of \SI{2}{\percent} into the isomeric state. Thus tagging of the ions in flight is essential to obtain a timing signal determining the optimum time for closing the Paul trap with an ion inside. Finally, a design for the setup of the TACTICa source is given which includes the possibility to load both recoil ions or ions from a laser ablation ion source into the Paul trap. The construction of the setup is ongoing.

\section{Acknowledgements}
	This work is supported by the Helmholtz Excellence Network ExNet020, Precision Physics, Fundamental Interactions and Structure of Matter (PRISMA+) from the Helmholtz Initiative and Networking Fund. Parts of this research were conducted using the supercomputer Mogon and/or advisory services offered by Johannes Gutenberg University Mainz (hpc.uni-mainz.de), which is a member of the AHRP (Alliance for High Performance Computing in Rhineland Palatinate, www.ahrp.info) and the Gauss Alliance e.V.

%
\section*{Conflict of interest}
The authors declare that they have no conflict of interest.



\end{document}